\documentclass[pra,aps,twocolumn,showpacs]{revtex4}

\usepackage{amsmath}
\usepackage{bm}
\usepackage{graphicx}

\begin{document}

\title{Dynamics of bubbles in a two-component Bose-Einstein	condensate}

\author{Kazuki Sasaki}
\author{Naoya Suzuki}
\author{Hiroki Saito}

\affiliation{
Department of Engineering Science, University of Electro-Communications,
Tokyo 182-8585, Japan
}

\date{\today}

\begin{abstract}
The dynamics of a phase-separated two-component Bose-Einstein condensate
are investigated, in which a bubble of one component moves through the
other component.
Numerical simulations of the Gross--Pitaevskii equation reveal a variety
of dynamics associated with the creation of quantized vortices.
In two dimensions, a circular bubble deforms into an ellipse and splits
into fragments with vortices, which undergo the Magnus effect.
The B\'enard--von K\'arm\'an vortex street is also generated.
In three dimensions, a spherical bubble deforms into toruses with vortex
rings.
When two rings are formed, they exhibit leapfrogging dynamics.
\end{abstract}


\maketitle

\section{Introduction}

A droplet of ink falling into water deforms from a sphere to a toroidal
shape due to the formation of a vortex ring~\cite{Thomson,Yajima}.
An air bubble rising in water exhibits complicated dynamics such as zigzag
and spiral motion~\cite{Haberman,Saffman}.
Such phenomena are caused by the interaction between the droplet (ink) or
the bubble (air) with the medium (water), where the former moves through
the latter.
The subject of the present paper is the dynamics of phase-separated
two-component superfluids in similar situations, that is, a bubble of one
component moving through the other component.

A variety of dynamical properties have been observed in two-component
Bose--Einstein condensates (BEC)~\cite{Myatt}.
Two immiscible components prepared in a spatially overlapped state
separate dynamically~\cite{Hall} and develop into a domain
structure~\cite{Miesner}.
A similar structure is observed in a miscible system with
counterflow~\cite{Hoefer}.
In rotating two-component BECs, a square vortex lattice has been
observed~\cite{Schweik}.
These dynamical properties of two-component BECs have been studied
theoretically~\cite{Chui,Kasamatsu1,Law,Ishino,Mueller,Kasamatsu2}.
Theoretical predictions have been made on various structures in
two-component BECs; for example, Skyrmions~\cite{Ruo}, solitary wave
complexes~\cite{Berloff}, and vortex bright solitons~\cite{KJHLaw}.
Various interface instabilities known in classical fluid mechanics are
predicted to emerge in two-component BECs with an interface, namely
the Rayleigh--Taylor instability~\cite{Sasaki,Gautam}, the
Kelvin--Helmholtz instability~\cite{Takeuchi,Suzuki}, and the
Richtmyer--Meshkov instability~\cite{Bezett}.
The behavior of two-component BECs strongly depends on the miscibility,
which is determined by the intra- and inter-component interactions.
Recently, these interactions in two-component systems have been
controlled using the Feshbach resonance~\cite{Thal}, and phase separation
dynamics have been studied in a controlled manner~\cite{Papp,Tojo}.

In the present paper, we investigate the dynamics of a bubble in a
phase-separated two-component BEC, where a small fraction of one component
moves through the other component.
We show that the system exhibits a rich variety of dynamics depending on
the parameters and dimensionality.
In two dimensions (2D), a circular bubble at rest deforms into an
ellipse as it accelerates and breaks into pieces with quantized
vortices.
For strong phase separation, a vortex street is formed in the wake of the
moving bubble, as in the B\'enard--von K\'arm\'an vortex street in a
single-component BEC~\cite{SasakiL}.
In three dimensions (3D), a spherical bubble deforms into a torus, as in
classical fluids~\cite{Thomson,Yajima}.
The significant difference from classical fluids is that the toroidal
bubble is accompanied by a quantized vortex ring.
When two or more rings are generated, they leapfrog each other.

This paper is organized as follows.
Section~\ref{s:formulate} provides a formulation of the problem.
Section~\ref{s:2ddyn} numerically demonstrates the dynamics of a 2D system
and Sec.~\ref{s:bubble} analyzes the deformation of a bubble.
Section~\ref{s:3d} performs full 3D numerical simulations.
Section~\ref{s:conc} gives conclusions to this study.

\section{Formulation of the problem}
\label{s:formulate}

We study the dynamics of a two-component BEC using the zero-temperature
mean-field theory.
The dynamics of the system are described by the Gross--Pitaevskii (GP)
equation given by
\begin{subequations} \label{GP}
\begin{eqnarray}
i \hbar \frac{\partial \psi_1}{\partial t} & = & \left( -\frac{\hbar^2}{2
m_1} \nabla^2 + V_1 \right) \psi_1 + g_{11} |\psi_1|^2 \psi_1 
\nonumber \\
& & + g_{12} |\psi_2|^2 \psi_1,
\label{GP1} \\
i \hbar \frac{\partial \psi_2}{\partial t} & = & \left( -\frac{\hbar^2}{2
m_2} \nabla^2 + V_2 \right) \psi_2 + g_{22} |\psi_2|^2 \psi_2
\nonumber \\
& & + g_{12} |\psi_1|^2 \psi_2,
\end{eqnarray}
\end{subequations}
where $\psi_j$ is the macroscopic wave function, $m_j$ is the atomic mass,
and $V_j$ is the external potential for the $j$th component ($j = 1, 2$).
The interaction parameters $g_{jj'}$ are defined as
\begin{equation}
g_{jj'} = 2 \pi \hbar^2 a_{jj'} (m_j^{-1} + m_{j'}^{-1}),
\end{equation}
where $a_{jj'}$ is the $s$-wave scattering length between the atoms in the
$j$th and $j'$th components.

We assume that the interaction parameters satisfy
\begin{equation}
g_{11} g_{22} > g_{12}^2,
\end{equation}
and therefore the two components are phase separated~\cite{Pethick}.
We also assume that the external potentials for both components are
initially absent, $V_1 = V_2 = 0$.
We consider an initial state in which a small fraction of component 2 is
localized and surrounded by component 1, and $\psi_1$ is uniform far from
component 2.
The ground state is therefore a circular (2D) or spherical (3D) ``bubble''
of component 2 located in the sea of component 1.
At $t = 0$, we start to exert a force $F$ on the bubble by applying a
potential $V_2 = F x$ to component 2.
The bubble then moves in the $-x$ direction.
This is done, e.g., by applying a magnetic field gradient to a system in
which the atoms in component 2 have a magnetic dipole moment in the
direction of the magnetic field while those in component 1 do not.

We numerically solve Eq.~(\ref{GP}) by the pseudo-spectral method.
The initial state is the ground state for $V_1 = V_2 = 0$ obtained by the
imaginary-time propagation method.
We add a small amount of white noise to the initial state to break the
numerically exact symmetry.
The boundary in the numerical calculation is set far from the relevant
region so that the periodic boundary condition does not affect the
results.
In the following analysis, we assume $m_1 = m_2 \equiv m$ and $g_{11} =
g_{22} \equiv g$ to reduce the number of parameters.

\section{Two-dimensional system}
\label{s:2d}

\subsection{Dynamics}
\label{s:2ddyn}

We first demonstrate the numerical results for the 2D system.
The wave function is assumed to have the form $\psi_j(\bm{r}, t) = \psi_{j
\perp}(x, y, t) \phi(z)$, where the dynamics of the normalized wave
function $\phi(z)$ are frozen.
Integrating the GP equation with respect to $z$, the effective interaction
strength is found to be $g_{jj'} \int |\phi|^4 dz$.
Hence, the healing length is $\xi_{\rm 2D} = \hbar / (m g n_{\rm 2D}
\int |\phi|^4 dz)^{1/2}$ and the sound velocity is $v_{\rm s}^{\rm 2D} =
(g n_{\rm 2D} \int |\phi|^4 dz / m)^{1/2}$, where $n_{\rm 2D}$ is the 2D
density $|\psi_{1 \perp}|^2$ far from the bubbles.
Normalizing length and time by $\xi_{\rm 2D}$ and $\xi_{\rm 2D} / v_{\rm
s}^{\rm 2D}$, the intra-component interaction parameter in the GP equation 
becomes unity and the relevant interaction parameter is only $g_{12} /
g$.

\begin{figure}[t]
\includegraphics[width=8.5cm]{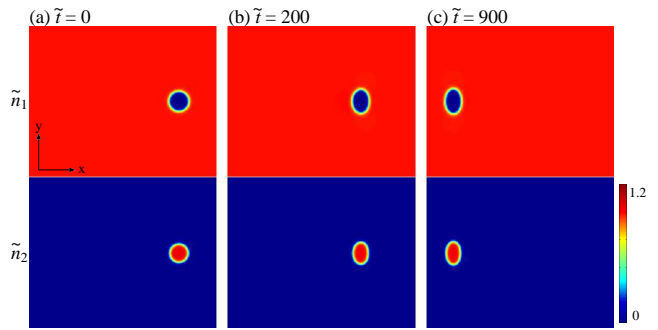}
\caption{
(Color online) Time evolution of the normalized density distribution
$\tilde n_j = |\psi_{j \perp}|^2 / n_{\rm 2D}$ for $g_{12} / g = 1.25$.
From $\tilde t = t v_{\rm s}^{\rm 2D} / \xi_{\rm 2D} = 0$ to $\tilde t =
200$, the force $\tilde F = F \xi_{\rm 2D}^2 / (\hbar v_{\rm s}^{\rm 2D})
= 1.58 \times 10^{-3}$ is exerted on component 2 in the $-x$ direction.
The amount of component 2 is $\int |\psi_2|^2 dx dy / (n_{\rm 2D}
\xi_{\rm 2D}^2) = 205$.
The field of view is $158 \times 126$ in units of $\xi_{\rm 2D}$.
}
\label{f:ellipse}
\end{figure}
Figure~\ref{f:ellipse} shows the time evolution of the density
distribution.
The initial state is a circular bubble as shown in Fig.~\ref{f:ellipse}
(a).
At $t = 0$, we apply a potential $V_2 = F x$ to component 2, and the
bubble is accelerated in the $-x$ direction.
We find that the bubble is deformed in an elliptic shape as shown in
Fig.~\ref{f:ellipse} (b). 
After the potential $V_2$ is switched off at $\tilde t = t v_{\rm s}^{\rm
2D} / \xi_{\rm 2D} = 200$, the bubble moves at a constant velocity 
keeping the elliptic shape as shown in Fig.~\ref{f:ellipse} (c).
Such elliptic deformation of a bubble is known in classical fluid
mechanics~\cite{Haberman} and will be analyzed in Sec.~\ref{s:bubble}.

\begin{figure}[t]
\includegraphics[width=8.5cm]{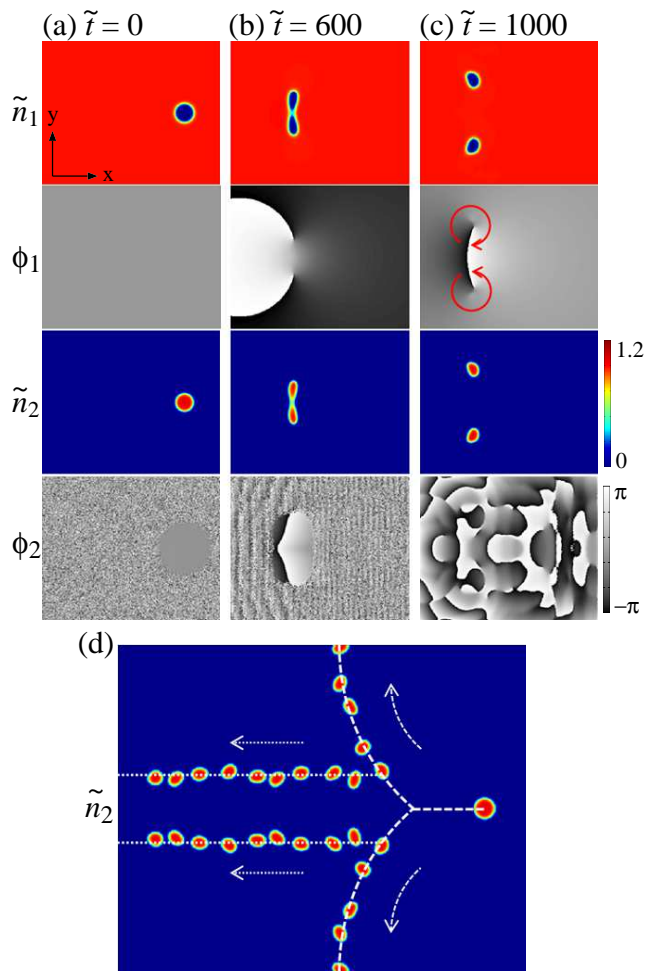}
\caption{
(Color online) (a)-(c) Time evolution of the density $\tilde n_j$ and
phase $\phi_j = {\rm arg} \psi_j$ profiles.
The force $\tilde F = 1.58 \times 10^{-3}$ is kept constant for $t > 0$.
Other parameters are the same as those in Fig.~\ref{f:ellipse}.
The arrows indicate the directions of circulation of the vortices.
(d) Stroboscopic trajectories of component 2, where the images are taken
at an interval of $\delta \tilde t = 1000$.
The force is constant for the images lying on the dashed lines, and the
force is switched off at $\tilde t = 1000$ for the images lying on the
dotted lines.
The field of view is $158 \times 126$ in (a)-(c) and $316 \times 253$ in
(d) in units of $\xi_{\rm 2D}$.
}
\label{f:split}
\end{figure}
Figures~\ref{f:split} (a)-\ref{f:split} (c) show the dynamics for the same
parameters as those in Fig.~\ref{f:ellipse}, where the force is exerted
for $t > 0$.
The bubble is first deformed elliptically as in Fig.~\ref{f:ellipse} (b)
and then splits into two bubbles [Figs.~\ref{f:split} (b) and
\ref{f:split} (c)].
We note that component 1 contains singly-quantized vortices with opposite
circulations at the split bubbles as indicated by the arrows in
Fig.~\ref{f:split} (c).
After the split, the two bubbles move away from each other as shown by the
dashed lines in Fig.~\ref{f:split} (d), and eventually move in the
directions perpendicular to the force.
This behavior is due to the Magnus force on the vortices given by
\begin{equation}
\bm{F}_{\rm M} = \rho \bm{\kappa} \times \bm{v},
\end{equation}
where $\rho$ is the mass density, $\bm{\kappa}$ is the vorticity, and
$\bm{v}$ is the velocity of the vortex.
In this case, $\rho \simeq m n_{\rm 2D}$ and $\kappa = 2 \pi \hbar / m$.
The velocity of the bubbles in the $\pm y$ directions is estimated by
equating the Magnus force in the $+x$ direction and the external force on
the bubble in the $-x$ direction as
\begin{equation} \label{magvel}
m n_{\rm 2D} \frac{2 \pi \hbar}{m} v_y = F \int |\psi_{2 \perp}|^2
dx dy,
\end{equation}
where the integration is taken within one of the bubbles.
For the parameters in Fig.~\ref{f:split}, the velocity is estimated from
Eq.~(\ref{magvel}) to be $v_y / v_{\rm s}^{2D} \simeq 0.026$, which is in
good agreement with the velocity $\simeq 0.026$ along the dashed line in
Fig.~\ref{f:split} (d).
If the external potential for component 2 is switched off after the bubble
splits, the two bubbles move in the $-x$ direction at a constant velocity
$\simeq \kappa / (2 \pi d)$ with a constant distance $d$ maintained
between the bubbles, as shown by the dotted lines in Fig.~\ref{f:split}
(d).
Such a two-component vortex is suitable for studying the Magnus effect,
since we can exert a force on the vortex in a controlled manner.

\begin{figure}[t]
\includegraphics[width=8.5cm]{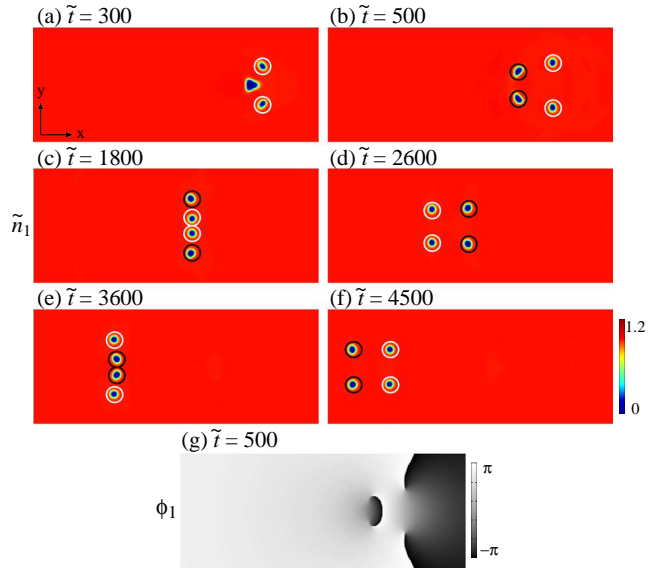}
\caption{
(Color online) (a)-(f) Time evolution of the density profile of component
1, where the force $\tilde F = 4.74 \times 10^{-3}$ is exerted on
component 2 from $\tilde t = 0$ to $\tilde t = 500$.
The other conditions are the same as those in Fig.~\ref{f:ellipse}.
The two vortex pairs are marked by black and white circles to follow their
motion.
(g) Phase profile of component 1 at $\tilde t = 500$.
The field of view in (a)-(g) is $316 \times 126$ in units of $\xi_{\rm
2D}$. 
}
\label{f:multi}
\end{figure}
Figures~\ref{f:multi} (a)-\ref{f:multi} (f) shows the time evolution of
the density profile of component 1, where the force is stronger than that
in Figs.~\ref{f:ellipse} and \ref{f:split}.
At $\tilde t = 300$, the bubble releases two fragments, at which point
component 1 has a vortex-antivortex pair [two white circles in
Fig.~\ref{f:multi} (a)].
The preceding bubble then splits into two, at which point a
vortex-antivortex pair is also contained in component 1 [black circles in
Fig.~\ref{f:multi} (b)].
After that, the pair marked by the white circles passes through that
marked by the black circles, and the two pairs leapfrog each other back
and forth as shown in Figs.~\ref{f:multi} (c)-\ref{f:multi} (f).

\begin{figure}[t]
\includegraphics[width=8.5cm]{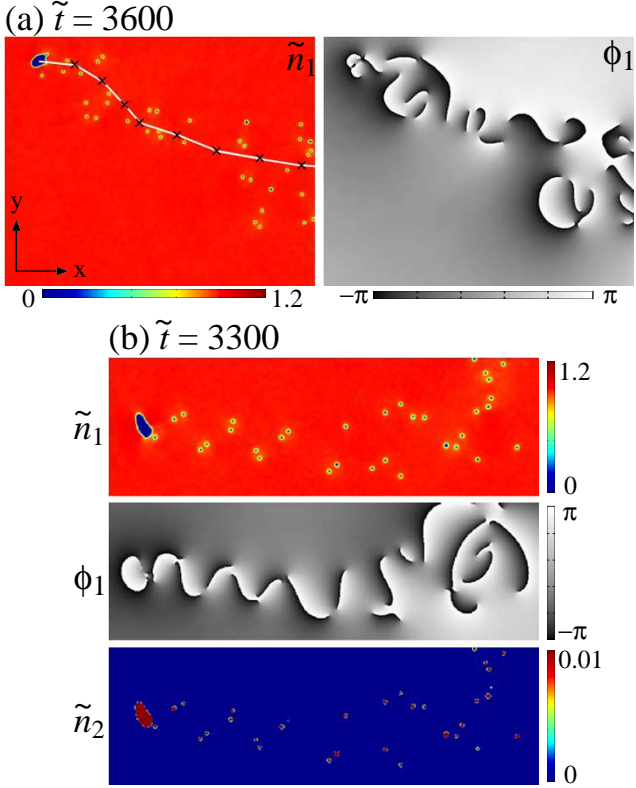}
\caption{
(Color online) Snapshots of the density and phase profiles for $g_{12} / g
= 3.3$.
The force on component 2 is $\tilde F = 6.96 \times 10^{-3}$ in (a)
and $\tilde F = 6.01 \times 10^{-3}$ in (b).
The black crosses in (a) indicate the positions of the bubble at an
interval of $\Delta \tilde t = 200$.
The field of view is $-869 \xi_{\rm 2D} < x < -474 \xi_{\rm 2D}$ and
$-158 \xi_{\rm 2D} < y < 158 \xi_{\rm 2D}$ in (a) and $-869 \xi_{\rm 2D} <
x < -474 \xi_{\rm 2D}$ and $-63 \xi_{\rm 2D} < y < 63 \xi_{\rm 2D}$ in
(b), where the initial spherical bubble is located at $x = y = 0$.
The amount of component 2 is the same as that in Fig.~\ref{f:ellipse}.
}
\label{f:karman}
\end{figure}
Figure~\ref{f:karman} shows the cases for strong force and strong phase
separation.
Since the interface tension is large for strong phase
separation~\cite{Barankov,Schae}, the bubble is hard to split as in
Fig.~\ref{f:split}.
Consequently, the bubble is accelerated beyond the critical velocity for
vortex creation and the vortices are shed in the wake as shown in
Fig.~\ref{f:karman}.
Figure~\ref{f:karman} (a) shows a typical behavior of the bubble, in which
the Magnus force bends the trajectory of the bubble while it contains
vortices and then the bubble moves in a winding fashion since the
vorticity contained in the bubble changes in time.
Occasionally, vortices are shed in pairs periodically as shown in
Fig.~\ref{f:karman} (b), which is reminiscent of the B\'enard--von
K\'arm\'an vortex street in a single-component BEC~\cite{SasakiL}.
Unlike vortex shedding in a single-component BEC, the bubble gradually
diminishes because the cores of the released vortices are occupied by
component 2 as shown in the bottom panel of Fig.~\ref{f:karman} (b).
The condition thus continuously changes and the vortex street generation
does not last long, since the parameter region for the vortex street
generation is narrow~\cite{SasakiL}.

\subsection{Analysis of bubble deformation}
\label{s:bubble}

To understand the elliptic deformation of the 2D bubble shown in
Fig.~\ref{f:ellipse}, we perform a simple analysis.

In this subsection, we assume that the thickness of the interface between
the two components is negligible due to the strong phase separation, and
the effect of the interface is expressed only by an interface tension
coefficient $\alpha$.
The wave function of the stationary state of component 1 outside the
bubble is written as $\psi_1(\bm{r}, t) = n_1^{1/2}(\bm{r}) \exp[i
	\theta_1(\bm{r}) - i \mu_1 t / \hbar]$, where we consider the problem in
the frame moving with the bubble.
Substituting this expression into the GP equation (\ref{GP}), we obtain
\begin{eqnarray}
\label{hydro1}
\nabla \cdot [n_1(\bm{r}) \bm{v}_1(\bm{r})] & = & 0, \\
\label{hydro2}
-\frac{\hbar^2}{2m} \frac{\nabla^2 \sqrt{n_1(\bm{r})}}{\sqrt{n_1(\bm{r})}}
+ \frac{1}{2} m v_1^2(\bm{r}) + g_{11} n_1(\bm{r}) & = & \mu_1,
\end{eqnarray}
where $\bm{v}_1 = \hbar \nabla \theta_1 / m$.
We also assume that the system is almost incompressible, i.e.,
$n_j(\bm{r}) = \bar{n}_j + \delta n_j(\bm{r})$ with $\delta n_j(\bm{r})
\ll \bar{n_j}$.
Equations~(\ref{hydro1}) and (\ref{hydro2}) then become
\begin{eqnarray}
\nabla \cdot \bm{v_1}(\bm{r}) & \simeq & 0, \\
\label{Bernoulli}
P_1(\bm{r}) + \frac{1}{2} \bar{n}_1 m v_1^2(\bm{r}) & \simeq & {\rm
const.},
\end{eqnarray}
where $P_1(\bm{r}) = g_{11} n_1^2(\bm{r}) / 2$ is the pressure.
Equation~(\ref{Bernoulli}) corresponds to Bernoulli's equation in
classical fluid mechanics.

The shape of the bubble is approximated to be an ellipse given by $x^2 /
a^2 + y^2 / b^2 = 1$, and we estimate $a$ and $b$ considering the pressure
inside and outside the bubble at $\bm{r}_a = (a, 0)$ and $\bm{r}_b = (b,
0)$.
Since $v_1 = 0$ at the stagnation point $\bm{r}_a$, Eq.~(\ref{Bernoulli})
gives
\begin{equation} \label{prel}
P_1(\bm{r}_a) = P_1(\bm{r}_b) + \frac{1}{2} \bar{n}_1 m v_1^2(\bm{r}_b).
\end{equation}
For inviscid and incompressible flow, the velocity on either side of an
elliptic obstacle is $v_1(\bm{r}_b) = v_0 (a + b) / a$~\cite{Lamb}, where
$v_0$ is the velocity at infinity.
Applying Laplace's formula~\cite{Landau} to the inner pressure at
$\bm{r}_a$ and $\bm{r}_b$, we have
\begin{equation} \label{laplace}
P_1(\bm{r}_a) + \alpha \frac{a}{b^2} = P_1(\bm{r}_b) + \alpha
\frac{b}{a^2},
\end{equation}
where $a / b^2$ and $b / a^2$ are the curvatures of an ellipse at
$\bm{r}_a$ and $\bm{r}_b$, respectively.
Using Eqs.~(\ref{prel}) and (\ref{laplace}), we obtain an expression that
estimates the oblateness of the bubble as
\begin{equation} \label{mainrel}
\frac{m \bar{n}_1 v_0^2}{2 \alpha} \sqrt{\frac{A}{\pi}} = \sqrt{\beta}
\frac{1 - \beta^3}{(1 + \beta)^2},
\end{equation}
where $A = \pi a b$ is the area of the ellipse and $\beta = a / b$.
The left-hand side of Eq.~(\ref{mainrel}) corresponds to the Weber number
in fluid mechanics.

\begin{figure}[t]
\includegraphics[width=9cm]{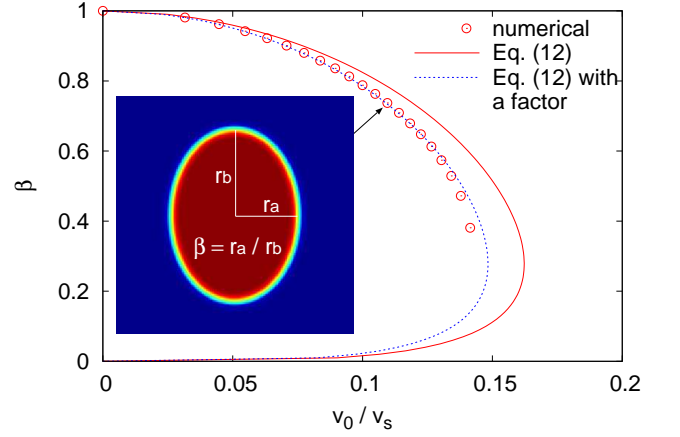}
\caption{
(Color online) Oblateness $\beta$ of an elliptic bubble as a function of
	the velocity $v_0$.
The solid curve shows Eq.~(\ref{mainrel}) with the interface tension
coefficient $\alpha$ in Eq.~(\ref{alpha}).
The dotted curve shows Eq.~(\ref{mainrel}) with the left-hand side
multiplied by 1.2.
The circles are numerically obtained from the GP equation.
The inset shows an example of the density distribution of component 2 for
a stationary state.
}
\label{f:analytic}
\end{figure}
We compare Eq.~(\ref{mainrel}) with the numerical calculation.
We numerically solve Eq.~(\ref{GP}) with $V_1 = V_2 = 0$ and $-i v_0
\partial_x \psi_j$ added to the right-hand side.
The imaginary time propagation of this equation gives the stationary
state in the frame moving in the $-x$ direction with velocity $v_0$.
The circles in Fig.~\ref{f:analytic} plot the ratio of the size of the
bubble in the minor axis to that in the major axis as a function of the
velocity.
The solid curve in Fig.~\ref{f:analytic} shows Eq.~(\ref{mainrel}) with the
interface tension coefficient~\cite{Barankov,Schae},
\begin{equation} \label{alpha}
\alpha = \frac{\hbar \bar{n}_1^{3/2}}{\sqrt{2m}} \sqrt{g_{12} - g}.
\end{equation}
We find that the solid curve deviates from the circles.
The deviation of the analytic result from the numerical result may be due
to the approximations and assumptions made in the analysis.
This deviation can be compensated if we modify Eq.~(\ref{mainrel}).
The dotted curve in Fig.~\ref{f:analytic} shows Eq.~(\ref{mainrel}) with
the left-hand side multiplied by 1.2, which is in good agreement with the
circles in Fig.~\ref{f:analytic}.

\section{Three dimensional system}
\label{s:3d}

In the numerical simulations in 3D, we normalize length and time by $\xi =
\hbar / (m g n_{\rm 3D})^{1/2}$ and $\xi / v_{\rm s}$, where $n_{\rm 3D}$
is the atom density far from the bubbles and $v_{\rm s} = (g n_{\rm 3D} /
m)^{1/2}$ is the sound velocity.

\begin{figure}[t]
\includegraphics[width=8.5cm]{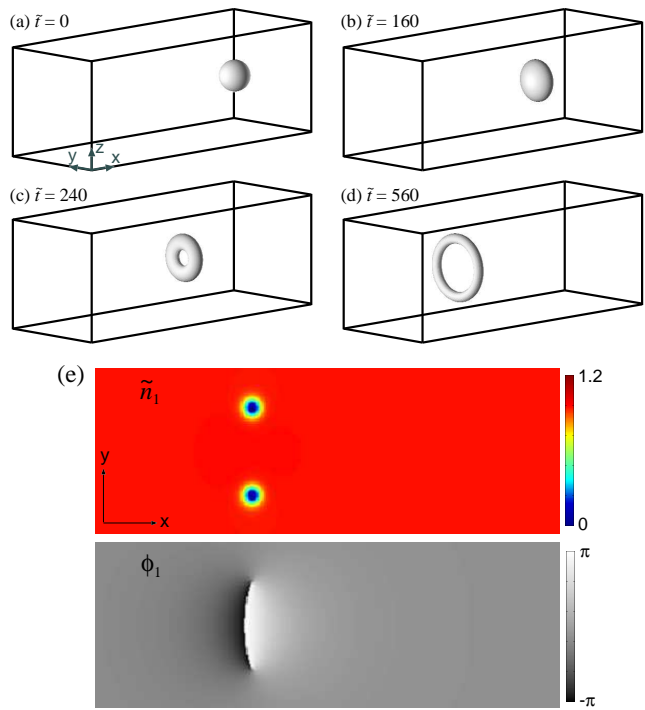}
\caption{
(Color online) (a)-(d) Time evolution of the isodensity surface of
	component 2 in the 3D simulation.
The size of the frame is $178 \times 63 \times 63$ in units of $\xi$.
(e) Density $\tilde n_1 = |\psi_1|^2 / n_{\rm 3D}$ and phase $\phi_1 =
{\rm arg} \psi_1$ profiles of component 1 on the $z = 0$ cross section at
$\tilde t = t v_{\rm s} / \xi = 560$.
The field of view is $178 \times 63$ in units of $\xi$.
The parameters are $g_{12} / g = 1.1$, $\tilde F = F \xi^2 / (\hbar
v_{\rm s}) = 3.2 \times 10^{-3}$, and $\int |\psi_2|^2 d\bm{r} / (n_{\rm
3D} \xi^3) = 3.2 \times 10^3$.
}
\label{f:ring}
\end{figure}
Figures~\ref{f:ring} (a)-\ref{f:ring} (d) show typical dynamics of a
bubble in 3D.
The initial state is a spherical bubble [Fig.~\ref{f:ring} (a)], which
deforms into a ellipsoidal shape as it is accelerated [Fig.~\ref{f:ring}
(b)].
The bubble then takes a toroidal shape [Fig.~\ref{f:ring} (c)], whose
radius increases in time [Fig.~\ref{f:ring} (d)].
Since the amount of component 2, $\int |\psi_2|^2 d\bm{r}$, is conserved,
the torus becomes thin as its radius increases.
Figure~\ref{f:ring} (e) shows the density and phase profiles on the cross
section of $z = 0$ at $\tilde t = t v_{\rm s} / \xi = 560$.
We find that component 1 contains a quantized vortex ring along the torus
of component 2.
If the force on component 2 is switched off after the ring is formed, the
ring propagates in the $-x$ direction at a constant velocity without
expansion of the radius.
The velocity roughly agrees with $v \simeq \hbar / (2 m R) \ln (8R /
a)$~\cite{Fetter}, where $R$ is the radius of the torus and $a$ is the
radius of the vortex core.

In contrast to a vortex ring in a single component
superfluid~\cite{Rayfield}, we can manipulate the vortex ring by applying
a potential to the torus of component 2 using a magnetic field gradient or
laser field.
We also note that a Skyrmion~\cite{Ruo} can be created if we imprint the
phase ${\rm arg}(z \pm i y)$ on the torus of component 2 using, e.g., the
Raman transition with Laguerre--Gaussian beams~\cite{Andersen}.

\begin{figure}[t]
\includegraphics[width=8.5cm]{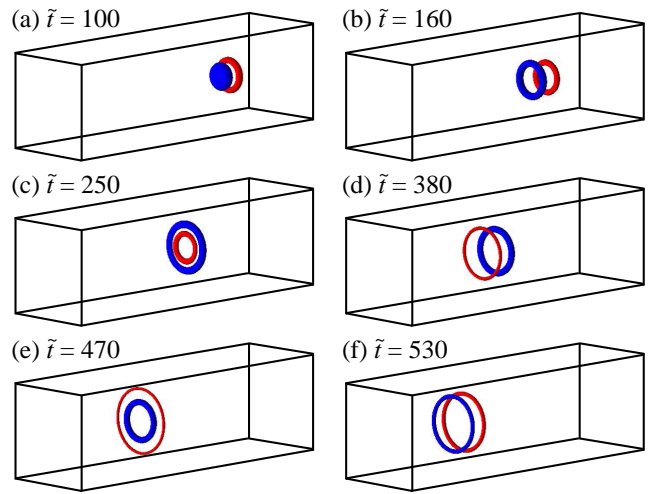}
\caption{
(Color online) Time evolution of the isodensity surface of
component 2 in the 3D simulation.
The force is $\tilde F = 9.5 \times 10^{-3}$ and the other parameters are
the same as those in Fig.~\ref{f:ring}.
The two objects are color coded for clarity.
The size of the frame is $221 \times 63 \times 63$ in units of $\xi$.
}
\label{f:leapfrog}
\end{figure}
When the force $F$ is strong, we can create multiple rings as shown in
Fig.~\ref{f:leapfrog}, where the two rings are color coded to distinguish
them.
The initial state is the same as that in Fig.~\ref{f:ring} (a).
The bubble first releases a ring backwards [Fig.~\ref{f:leapfrog} (a)]
and then the front bubble also deforms into a ring [Fig.~\ref{f:leapfrog}
(b)], resulting in double vortex rings with the same circulation.
After that, the rear ring passes through the front ring
[Figs.~\ref{f:leapfrog} (c) and \ref{f:leapfrog} (d)], and this overtaking
is repeated [Figs.~\ref{f:leapfrog} (d) and \ref{f:leapfrog} (f)], which
is the 3D version of the behavior in Fig.~\ref{f:multi}.
Such a leapfrogging behavior of vortex rings was first predicted in
Ref.~\cite{Helmholtz}.

\section{Conclusions}
\label{s:conc}

We have investigated the dynamics of phase-separated two-component BECs,
in which a ``bubble'' of one component propagates in the other component.
We studied 2D systems in Sec.~\ref{s:2d}.
When the velocity of the bubble is small, the circular bubble deforms into
an elliptic shape, which travels with a constant velocity if the force is
switched off (Fig.~\ref{f:ellipse}).
The elliptic deformation of the bubble was analyzed in
Sec.~\ref{s:bubble}.
For a large velocity, the bubble of component 2 splits into two or more
fragments, where component 1 contains quantized vortices
(Figs.~\ref{f:split} and \ref{f:multi}).
The trajectories of bubbles containing vortices are then affected by the
Magnus force [Fig.~\ref{f:split} (d)].
When the force and phase separation are strong, vortex shedding occurs
instead of the split.
The bubble drifts due to the Magnus force and sometimes generates the
B\'enard--von K\'arm\'an vortex street (Fig.~\ref{f:karman}).
For the 3D system studied in Sec.~\ref{s:3d}, we found that the spherical
bubble changes to a toroidal shape, where component 1 contains a quantized
vortex ring (Fig.~\ref{f:ring}).
When two vortex pairs or two vortex rings are created, they exhibit a
leapfrog behavior (Figs.~\ref{f:multi} and \ref{f:leapfrog}).

We have thus shown that bubbles in two-component BECs exhibit a rich
variety of phenomena.
When the bubble of component 2 splits in 2D or becomes a torus in 3D,
quantized vortices are generated in component 1, whose cores are occupied
by component 2.
We can therefore exert a force on the vortices in component 1 in a
controlled manner through the force on component 2, which may allow
manipulation of vortices in a BEC.

\begin{acknowledgments}  
We thank S. Tanaka for his participation in the early stages of this
work.
This work was supported by the Ministry of Education, Culture, Sports,
Science and Technology of Japan (Grants-in-Aid for Scientific Research,
No.\ 20540388 and No.\ 22340116).
\end{acknowledgments}

\end{document}